\documentclass[12pt]{iopart}
\usepackage[utf8]{inputenc}
\usepackage[compatibility=false]{caption}
\usepackage{subcaption}
\usepackage{cite}
\usepackage{graphicx}
\usepackage[colorlinks=true, allcolors=blue]{hyperref}
\usepackage{xcolor}

\begin{document}

\title{Impact of periodic thermal driving on heat fluctuations in a harmonic system}

\author{Felipe P Abreu$^1$, Welles A M Morgado$^1$}

\address{$^1$Pontifícia Universidade Católica do Rio de Janeiro (PUC-Rio), \\
Rua Marquês de São Vicente, 225 – Gávea, Rio de Janeiro, RJ, 22451-900, Brazil}
\ead{\mailto{fabreu@aluno.puc-rio.br}, \mailto{welles@puc-rio.br}}
\vspace{10pt}

\begin{abstract}
The thermodynamics of mesoscopic systems driven by time-varying temperatures is crucial for understanding biological systems, designing nanoscale engines, and performing micro-particle cooling. In this work, we analyze an underdamped Brownian particle in a harmonic trap under a sinusoidal thermal protocol. Through analytical methods and numerical simulations, we analyze the system's dynamics and heat statistics. We report the emergence of resonant position-velocity correlations and a non-Gaussian, asymmetric heat distribution consistent with the Fluctuation Theorem. We demonstrate that inertia is a key parameter, damping the system's response and slowing its relaxation to a periodic non-equilibrium steady state. Our results show that oscillatory thermal driving is a powerful tool for controlling nanoscale energy flow.
\end{abstract}

%
\vspace{2pc}
\maketitle
%
%

\section{Introduction}
In mesoscopic systems, energy exchanges are inherently stochastic. Thermodynamic quantities, such as work and heat, exhibit fluctuations and are therefore described within the framework of stochastic thermodynamics~\cite{Seifert2012,Peliti2021,Sekimoto2010}. In this framework, variable temperature processes emerge and play a fundamental role in many contexts, \textit{e.g.}, Brownian engines, micro-particle cooling, and biological systems~\cite{Martinez2016,Martinez_2017,Blickle2012}. 

To describe the motion of particles subjected to random forces in colloidal environments, a theoretical framework is provided by the Langevin and Fokker–Planck equations~\cite{Van_Kampen_2007,Sekimoto2010}. These equations are employed in two distinct regimes, namely the underdamped and overdamped limits, each capturing different features of the stochastic behavior. The underdamped regime ($m\neq 0$) accounts for inertia and preserves correlations between the particle’s position and velocity in non-equilibrium situations, whereas the overdamped regime ($m=0$) neglects inertia, decouples position from velocity, and is appropriate for systems with fast relaxation times. While both regimes have been extensively studied from a dynamical perspective~\cite{Safdari_2017,Tothova_2022,Huang_2011,Bodrova_2016,Naze_2022,Pebeu_2018}, their thermodynamic implications — particularly the role of inertia in energy exchange, dissipation, and system performance — remain less understood~\cite{Paraguassu_2023,Proesmans_2016,Proesmans_2017,Goswami_2019,Gomez_Solano_2024,Paraguass2025}.

In stochastic thermodynamics, heat exchange between a Brownian particle and its thermal environment is always present, whereas work is not necessarily involved~\cite{Paraguassu_2021}. Work is the ordered energy exchange between a system and an external agent in a controlled or directed manner. Therefore, for work to be performed, the Brownian particle must interact with an external agent, an external system~\cite{Sekimoto2010}.  In contrast, heat is the spontaneous and disordered energy exchange with the thermal bath, and occurs inherently in such systems~\cite{Sekimoto2010}. Despite their fundamental role, the energetics of Brownian systems cannot be directly accessed — typical experiments track the particle's trajectory over time~\cite{Roldan2014,Li2010}. To extract thermodynamic quantities, one must infer them from the statistical properties of the trajectories~\cite{Ciliberto2017}. Consequently, the theoretical framework is essential for interpreting empirical data and characterizing the underlying thermodynamic processes.

Focusing on optically trapped Brownian particles, new experimental methods are already being implemented to mimic effective, and thereby controllable, thermal baths~\cite{Pires,Raizen,Chupeau,Martinez2015,Laan_2021,Delic}.  Such controlled manipulation opens avenues for exploring a broad class of Brownian engines. Consequently, characterizing the thermodynamic properties of these non-isothermal processes is of considerable relevance for both fundamental studies in non-equilibrium systems and applications to nanoscale engines.

Motivated by these considerations, in this work we investigate how a temperature-time-dependent protocol affects the dynamics and thermodynamics of an underdamped Brownian particle in a harmonic potential. Specifically, we calculate analytical expressions for the system’s behavior under a generic thermal protocol and then evaluate these results for the case of a sinusoidal thermal protocol. Combining analytical methods with numerical simulations, we compute both the Pearson correlation coefficient between position and velocity and the heat distribution. For the Pearson correlation, we found a non-vanishing periodic correlation with resonant phenomena, while the heat distribution is highly sensitive to parameter control and is in accordance with the Fluctuation Theorem. 

This paper is organized as follows: In Section~\ref{Brown_dyn}, we describe the dynamics of the underdamped Brownian particle subjected to periodic thermal driving in a harmonic potential. In Section~\ref{Dynamics} we analyze the dynamics of the Pearson correlation coefficient. In Section~\ref{Heat_fluctuations} we analyze the probability distribution of the heat, derive the characteristic function, and calculate the moments of the distribution. We finish in Section~\ref{Conclusion} with the discussion of the results and conclusions. 

\section{Brownian dynamics with arbitrary temperature protocol}
\label{Brown_dyn}
We consider an underdamped Brownian particle model, with a temperature-time-dependent protocol $T(t)$. The Langevin equations are
\begin{eqnarray}
    \label{LE}
    \dot x(t) = v(t), \quad
    m \dot v(t) = -\gamma v(t) -\partial_x V(x(t))+\sqrt{2\gamma k_B T(t)}\,\xi(t), 
\end{eqnarray}
where $m, \gamma, k_B$ and $V(x(t))$ are the mass, damping coefficient, Boltzmann constant, and potential energy, respectively. $\xi(t)$ is Gaussian white noise with a zero mean and delta correlated, \textit{i.e.}, $\langle \xi(t)\rangle = 0$ and $\langle \xi(t)\xi(t')\rangle = \delta (t-t')$. Throughout this paper, we consider a harmonic potential $V(x) = kx^2/2$, where $k$ is an effective spring constant — for a Brownian particle trapped by an optical tweezer, $k$ is the  stiffness of the trap and depends linearly on the optical power~\cite{Felipe_2023}. This potential implies that Eq.\eref{LE} is linear, and since the statistics of $x(t)$ and $v(t)$ depend linearly on the Gaussian noise, the statistics of $x(t)$ and $v(t)$ are also Gaussian; therefore, the quantities of greater interest — which dictate the probability distribution — are the first and second moments. For a generic temperature-time-dependent protocol, these moments are given by
\begin{eqnarray}
    \langle x(t) \rangle &=&  \langle v(t) \rangle =0 \\
    \label{x2_med}
    \sigma_x^2(t) &=& \frac{k_B T_0 e^{-\beta t}}{k}\left[ 1 + \frac{\beta}{2\omega_1}\sin(2\omega_1 t) + \frac{\beta^2}{2\omega_1^2}\sin ^2(\omega_1 t) \right]\\ && + \frac{2\beta k_B}{m\omega_1^2}\int_0^t T(s) e^{-\beta  (t-s)} \sin ^2[\omega_1(s-t)] \,  d s \nonumber\\
    \label{v2_med}    
    \sigma_v^2(t) &=& \frac{k_B T_0 e^{-\beta t}}{m}\left[ 1 -\frac{\beta}{2 \omega_1}\sin (2\omega_1 t) + \frac{\beta ^2}{2\omega_1^2}\sin ^2(\omega_1 t) \right] \\ &&+ \frac{\beta k_B}{2m\omega_1^2}\int_0^t T(s) e^{-\beta(t-s)} \{\,\beta  \sin [\omega_1 (s-t)] + 2\omega_1 \cos [\omega_1 (s-t)]\,\}^2 \, d s \nonumber \\
    \label{xv_med}
    \langle x(t)v(t) \rangle &=& -\frac{\beta  k_B T_0 e^{-\beta t} \sin ^2(\omega_1 t)}{m \omega_1^2} \\ &&+ \frac{\beta k_B}{m\omega_1^2}\int_0^t  T(s)e^{-\beta(t-s)} \left\{\omega_1 \sin [2 \omega_1(t-s)]- \beta\sin ^2[\omega_1 (s-t)]\right\} d s. \nonumber
\end{eqnarray}
where $\beta=\gamma/m$, $\omega_0^2=k/m$ and $\omega_1 = \sqrt{\omega_0^2 - \beta^2/4}$ (See \ref{moments_derivation} for details). The foregoing equations assume that the system is initially at equilibrium with temperature $T_0$, \textit{i.e.}, the initial conditions obey $\langle x_0 \rangle = \langle v_0 \rangle = 0 $, $\langle x^2_0 \rangle = k_BT_0/k$ and $\langle v^2_0 \rangle = k_BT_0/m$. In addition, it is assumed $\omega_1>0$; if it is imaginary, the equations must be rewritten in terms of hyperbolic functions, using the relations $\cos(ix) = \cosh(x)$ and $\sin(ix) = i\sin(x)$.

Despite the calculations presented here being applicable to a generic temperature-time-dependent protocol, we focus the analysis on a specific case, a sinusoidal thermal protocol
\begin{equation}
    \label{Ts}
    T(t) = T_0 + T \sin (\Omega \,t),
\end{equation}
where $T$ defines the amplitude of the oscillations and $\Omega$ is the driving frequency. Such protocols have recently been suggested as a possible model for mechanical excitation and heat conduction in hydrocarbons~\cite{Chen2025}.

\section{Dynamics}
\label{Dynamics}

We analyze the Pearson correlation coefficient between displacement and velocity
\begin{eqnarray}
    \label{rho_t}
    \rho_{xv}(t) = \frac{\langle x(t)v(t) \rangle - \langle x(t) \rangle\langle v(t) \rangle}{\sigma_{v}(t)\sigma_{x}(t)}.
\end{eqnarray}
An important property of the Pearson correlation coefficient is $|\rho_{xv}(t)|\leq 1$. Moreover, if its absolute value is equal to one, $|\rho_{xv}(t)| = 1$, then there exist constants a and b such that v = ax + b with unit probability, \textit{i.e.} v(t) depends linearly on x(t)~\cite{Petruccione_2007}.

\begin{figure}[!ht]
\centering
\includegraphics[width=0.45\textwidth]{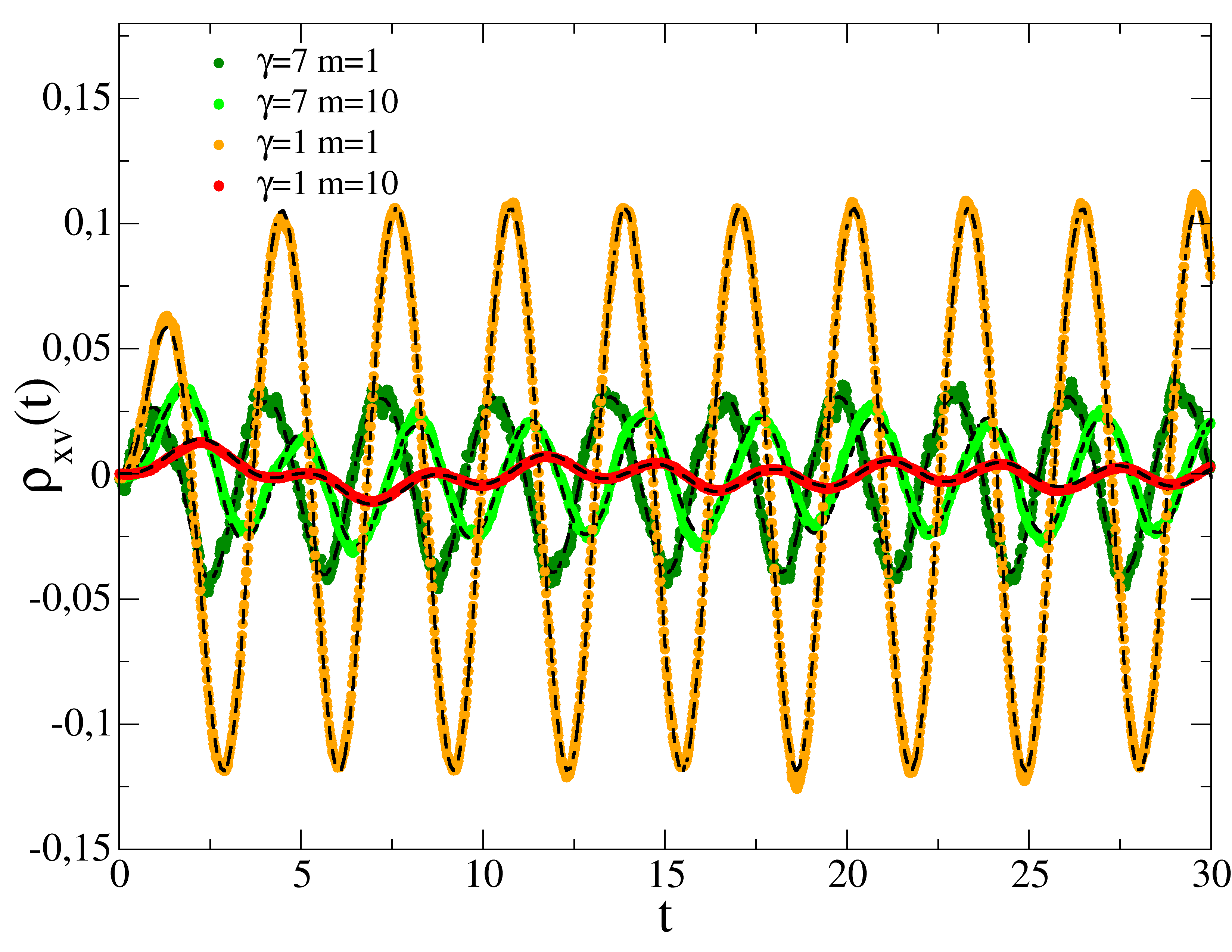} 
\caption{\label{fig:Pearson_correlation} Pearson correlation coefficient between the velocity and displacement for the Brownian particle under the sinusoidal-temperature protocol. Filled symbols correspond to data obtained from Brownian simulations with $k=1$, $T_0=20$, $T=5$ and $\Omega=2$, while dashed lines are the results expected from theory, \textit{i.e.}, Eq.\eref{rho_t}.}
\end{figure}

For a system with constant temperature, \textit{i.e.}, in thermal equilibrium, it is well known that $\rho_{xv}(t)$ is null. However, with a sinusoidal temperature protocol, \textit{i.e.}, modulation of noise, a residual periodic correlation appears, as can be seen in Fig.~\ref{fig:Pearson_correlation}.
Due to the time-dependent temperature $T(t)$, the naively overdamped approximation ($ m/\gamma \rightarrow 0$) fails to describe the long-time behavior ($t \gg m/\gamma$). The oscillatory temperature protocol imposes a rapid, periodic driving at frequency $\Omega$ that prevents the velocity from fully relaxing before $T(t)$ changes again. Physically, the small but non-zero value of $\rho_{xv}(t)$ is a signature of the residual inertial effects in a driven steady state. As the particle heats and cools (\textit{i.e.}, $T(t)$ oscillates) energy is periodically injected into the system, sustaining inertial effects.

\begin{figure}[ht!]
  \centering
  \begin{subfigure}[b]{0.45\linewidth}
    \subcaption{}         
    \includegraphics[width=\linewidth]{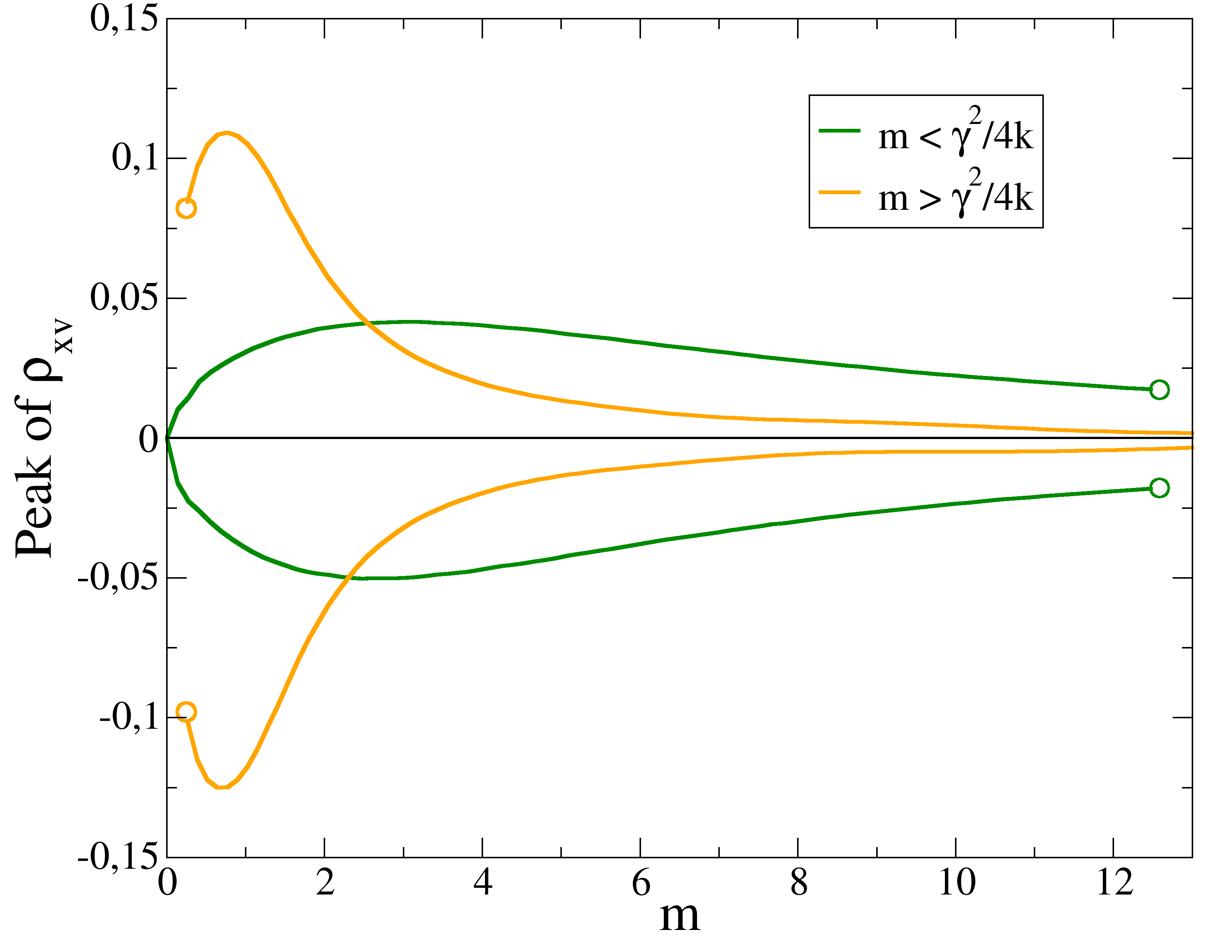}
    \label{fig:Peak_correlation}
  \end{subfigure}
  \begin{subfigure}[b]{0.47\linewidth}
    \subcaption{}
    \includegraphics[width=\linewidth]{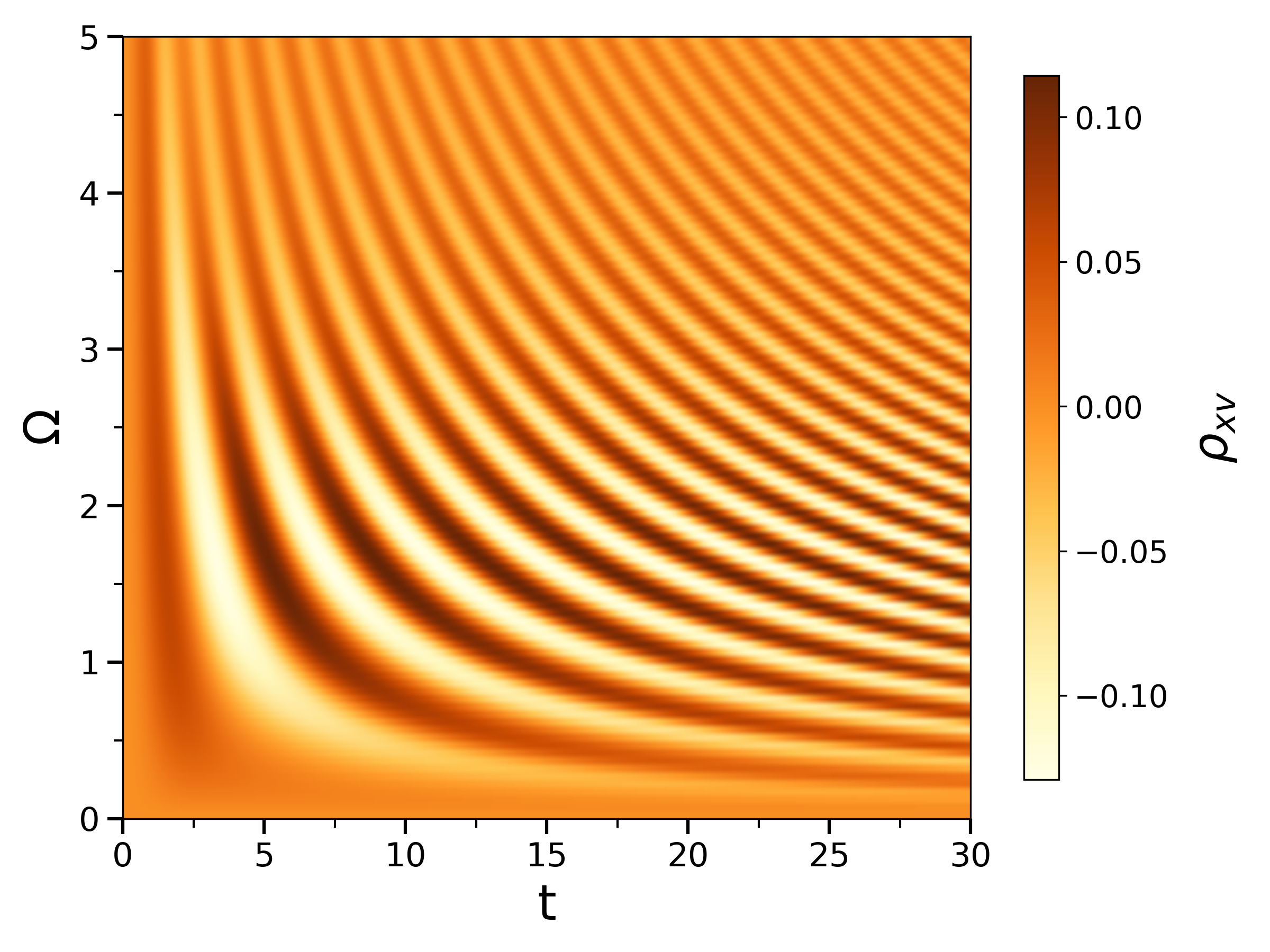}
    \label{fig:Heatmap_correlation}
  \end{subfigure}
  \caption{(a)  Peak of the Pearson correlation coefficient as function of the mass $m$. The empty circles indicate the critical damping points.  Here is considered $\Omega=2$, $\gamma = 7$ for the green line and $\gamma = 1$ for the orange line. (b) Heat map of the Pearson correlation with $m=1$. In both graphs is adopted $k= 1$, $T_0=20$, $T=5$.}
  \label{fig:Correlation_amplitude}
\end{figure}

As  displayed in Fig.~\ref{fig:Correlation_amplitude} the set of mass $m$, effective spring constant $k$, and friction coefficient $\gamma$ has a substantial influence on the correlation $\rho_{xv}(t)$. 
Numerical results in Fig.~\ref{fig:Peak_correlation} reveal the existence of two distinct regimes, set by a critical mass $m_c = \gamma^2/4k$. For $m>m_c$, $\omega_1$ is real, inertia dominates over friction, and the dynamics exhibits an oscillatory relaxation. For $m<m_c$, $\omega_1$  is purely imaginary, friction dominates over inertia, and the dynamics relaxes exponentially. At $m=m_c$ the system reaches the critical damping point, marking the boundary between these two regimes. At this precise mass, the system's response to any perturbation changes its fundamental character. This transition marks a fundamental change in the system's dynamical behavior. Moreover, in both regimes a peak is observed, indicating an optimal response for a given mass $m$ — a clear manifestation of a resonant phenomenon. On the other hand, as $m \rightarrow \infty$, the position-velocity correlation is significantly reduced and tends to zero.

One might be induced to think that the drive frequency $\Omega$ only influences the oscillation frequency on the Pearson correlation $\rho_{xv}(t)$. However, as displayed in Fig.~\ref{fig:Heatmap_correlation}, the amplitude of the correlation oscillations (\textit{i.e.}, the contrast between the dark and bright bands) varies with $\Omega$. The strongest correlations occur for driving frequencies in the range $\Omega\in [1,2]$, indicating the presence of another resonance phenomenon.

\section{Heat fluctuations}
\label{Heat_fluctuations}
Since the parameters of the system do not change, there is no work done upon it. So, from the first law of stochastic thermodynamics, the stochastic heat — which is a functional of the trajectory — is equal to the variation in the internal energy $Q[\,\vec{x}(t)\,] = \Delta K + \Delta V$. For this reason, the stochastic thermodynamics of the system can be characterized solely by the heat, and therefore we focus on it.

The stochastic heat is the competition between the noise of the bath and the drift force~\cite{Sekimoto2010}. By the underdamped dynamics in Eq.\eref{LE}, we can construct the stochastic heat functional 
\begin{equation}
    \label{stc_Q}
    Q[\vec{x}(t)] = \int_{x_0}^x \left(-\gamma v(t) + \xi(t)\right)\circ dx(t) = \frac{1}{2}m[v^2(t) - v_0^2] +  \frac{1}{2}k[x^2(t)-x_0^2] .
\end{equation}
Stratonovich-type product $\circ$ is adopted in the foregoing equation — as in macroscopic thermodynamics, $dQ>0$ assigns heat received by the Brownian particle, and $dQ<0$ assigns heat released to the thermal bath. Despite having a time dependency in the temperature, the heat functional is just the sum of the difference in the kinetic energy and the potential energy.

\subsection{Heat distribution}

\begin{figure}[!ht]
\centering
\includegraphics[width=0.5\textwidth]{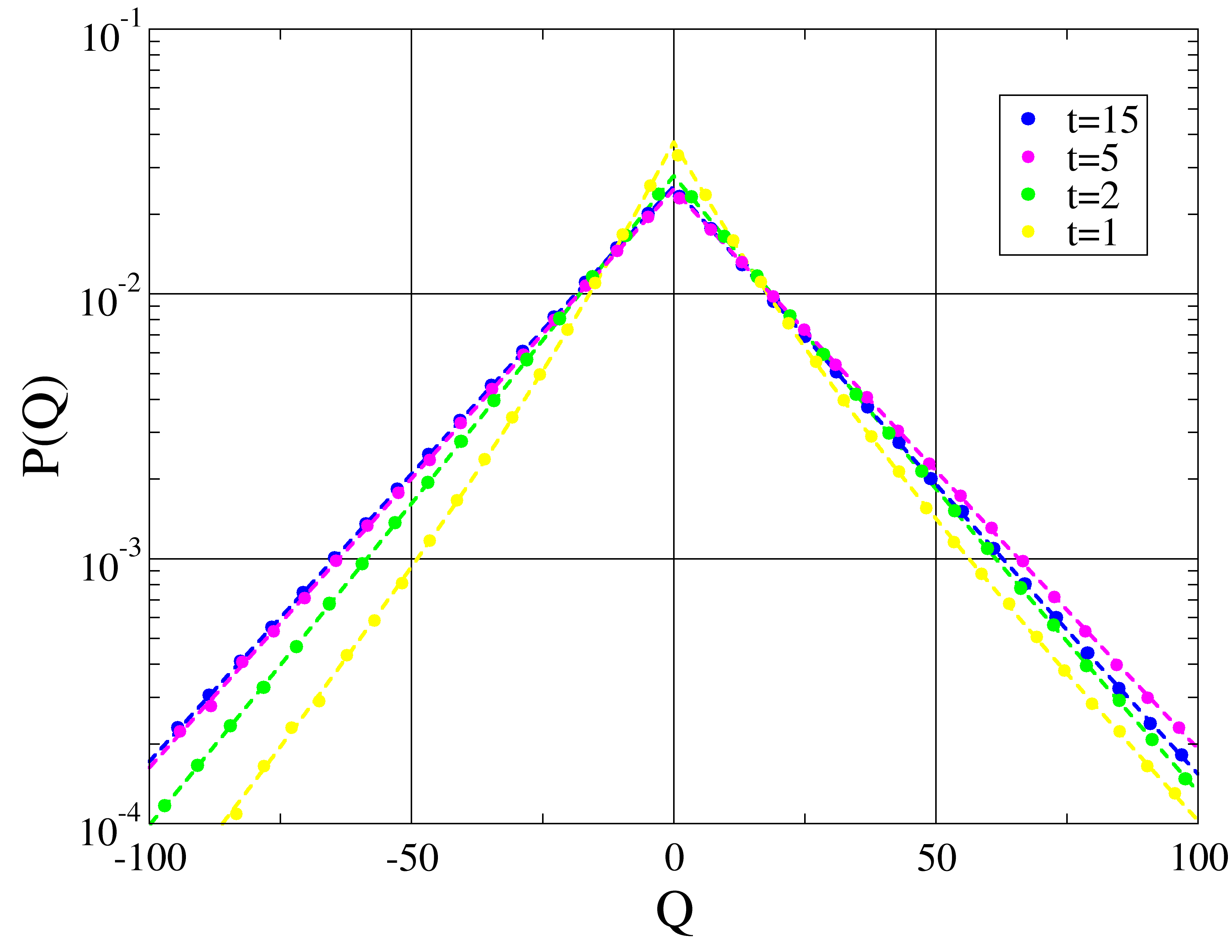} 
\caption{\label{fig:Heat_dist} Heat distribution as function of time. Filled symbols correspond to data obtained from Brownian simulations with $m=\gamma =k= 1$, $T_0=20$, $T=5$ and $\Omega=2$, and the dashed lines are the results expected from theory, \textit{i.e.}, Eq.\eref{P_Q} numerically integrated.}
\end{figure}

The random values of the heat $Q$ are given by the trajectory-dependent term $Q[\vec{x}(t)]$. Thereby, the \textit{probability density function} (PDF) for the heat is obtained by averaging over the trajectories 
\begin{eqnarray}
    \label{P_Q}
    P(Q) =  \langle \delta(Q-Q[\vec{x}(t)])\rangle = \frac{1}{2\pi}\int d \lambda\,  e^{i\lambda Q} \langle e^{-i \lambda Q[\vec{x}(t)]} \rangle.
\end{eqnarray}
The foregoing equation can be rewritten in terms of the characteristic function $Z(\lambda)$ ~\cite{Paraguassu2022}
 \begin{eqnarray}
    \label{Z_func}
    Z(\lambda) =   \langle e^{-i \lambda Q[\vec{x}(t)]} \rangle =  \int dx \int dv \int dx_0 \int dv_0 ~ P(x,v,x_0,v_0) e^{-i\lambda Q[\vec{x}(t)]} .
\end{eqnarray}
Since the integrals in Eq.\eref{Z_func} are convergent for any generic smooth and continuous temperature-time-dependent protocol, $Z(\lambda)$ contains the same amount of information as $P(Q)$. On the other hand, the integral in Eq.\eref{P_Q} cannot be written in terms of conventional functions, \textit{i.e.}, $P(Q)$ cannot be calculated analytically. For this reason, the heat PDF is numerically obtained, and the characteristic function is calculated analytically, which yields 
\begin{eqnarray}
    \label{Z_lambda}
    Z(\lambda) = \frac{1}{\sqrt{\det(\mathbf{\Psi}+\mathbf{I}_4)}}
\end{eqnarray}
where $\mathbf{I}_4$ is the identity matrix and $\mathbf{\Psi}$ is given by
\begin{eqnarray}
    \mathbf{\Psi} = i\lambda\left[ \begin{array}{cccc}
        k\, \sigma_x^2(t) & k\,\langle x(t)x_0\rangle & k\,\langle x(t)v(t)\rangle & k\,\langle x(t)v_0\rangle \\ -k\,\langle x(t)x_0\rangle & -k\,\sigma_{x_0}^2 & -k\,\langle v(t)x_0\rangle & 0 \\ m\,\langle x(t)v(t)\rangle & m\,\langle v(t)x_0\rangle & m\,\sigma_v^2(t) & m\,\langle v(t)v_0\rangle \\ -m\,\langle x(t)v_0\rangle & 0 & -m\,\langle v(t)v_0\rangle & -m\,\sigma_{v_0}^2
    \end{array}\right],     
\end{eqnarray}
where its elements are
\begin{eqnarray}
    \langle x_0v_0\rangle &=& 0 \\
    \sigma_{x_0}^2 &=& \frac{k_BT_0}{k} \\
    \sigma_{v_0}^2 &=& \frac{k_BT_0}{m} \\
    \langle x(t)x_0\rangle &=& \frac{k_B T_0 e^{-\beta  t/2}}{2 k \omega_1}[2 \omega_1 \cos(\omega_1 t)+\beta  \sin(\omega_1 t)] \\
    \langle x(t)v_0\rangle &=& \frac{k_B T_0 e^{-\beta  t/2}}{m \omega_1}\sin(\omega_1 t) \\ 
    \langle v(t)x_0\rangle &=& -\frac{k_B T_0 e^{-\beta  t/2}}{m \omega_1}\sin(\omega_1 t)\\   
    \langle v(t)v_0\rangle &=& \frac{k_B T_0 e^{-\beta  t/2}}{2 m \omega_1}[2 \omega_1 \cos(\omega_1 t)-\beta  \sin(\omega_1 t)].
\end{eqnarray}
(See \ref{characteristic_func} for details). As one can see in Fig.~\ref{fig:Heat_dist}, our numerical results present strong agreement with what is expected from the theory. 
Remarkably, we notice that the heat distribution is asymmetric around its peak. This asymmetry violates detailed balance, suggesting that the system is far from equilibrium. Moreover, Fig.~\ref{fig:Heat_dist} implies that the heat distribution curves are essentially linear in the semi-log scale, \textit{i.e.}, $P(Q) \propto e^{\beta_i Q}$. Since the slopes for $Q>0$ and $Q<0$ are different, there are two different decay constants, such that
\begin{eqnarray}
    P(Q) \propto \left\{ \begin{array}{rcl}
 e^{-\beta_+ Q} & \mbox{for}
& Q>0 \\ e^{-\beta_- |Q|} & \mbox{for} & Q<0. 
\end{array}\right.
\end{eqnarray}
This asymmetric exponential behavior is a direct manifestation of the heat fluctuation relation for a Brownian particle and a nonstationary bath~\cite{Solano_2011,Evans_2002,Koushik_2019}
\begin{eqnarray}
    \label{Heat_FT}
    \frac{P(Q)}{P(-Q)} = e^{\Delta\beta Q},    
\end{eqnarray}
where $\Delta\beta$ is related to the degree of entropy production of the system.  Thereby, the asymmetry observed is a quantitative feature predicted by non-equilibrium statistical mechanics.

\subsection{Statistical moments}
\begin{figure}[ht!]
  \centering
  \begin{subfigure}[b]{0.48\linewidth}
    \subcaption{}         
    \includegraphics[width=\linewidth]{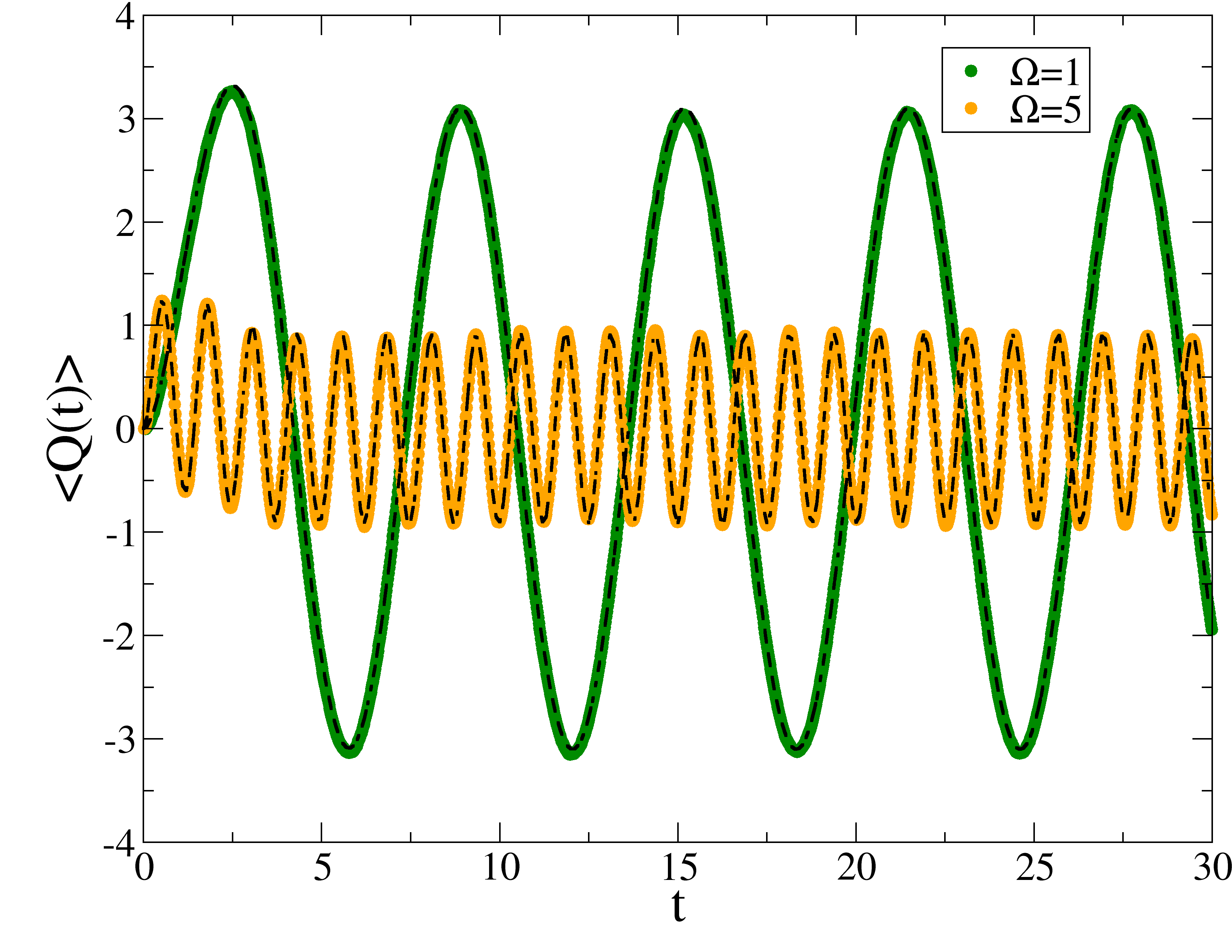}
    \label{fig:Heat_avg_m1}
  \end{subfigure}
  \begin{subfigure}[b]{0.48\linewidth}
    \subcaption{}
    \includegraphics[width=\linewidth]{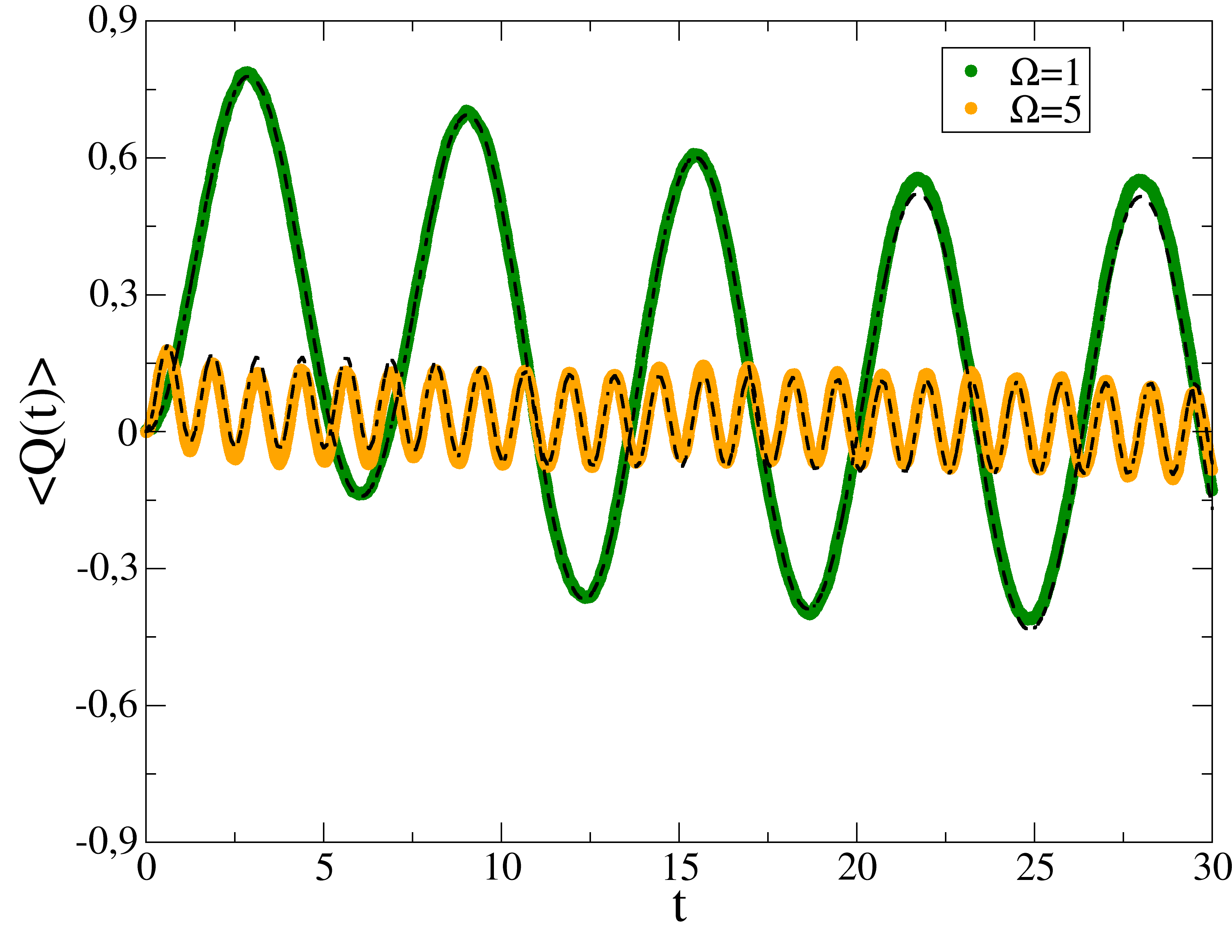}
    \label{fig:Heat_avg_m10}
  \end{subfigure}
  \caption{ Mean heat as function of time for the Brownian particle under the sinusoidal-temperature protocol for (a) m=1 and (b) m=10. In both graphs, filled symbols correspond to data obtained from Brownian simulations with $\gamma =k= 1$, $T_0=20$ and $T=5$. Dashed lines are the results expected from theory, \textit{i.e.}, Eq.\eref{Q_med}.}
  \label{fig:Heat_avg}
\end{figure}

The mean heat can be easily obtained by averaging Eq.\eref{stc_Q} over the trajectories, which yields
\begin{eqnarray}
    \label{Q_med}
    \langle Q(t) \rangle 
    &=& k_B T_0 \left[-1 +e^{-\beta t} \left(\frac{\beta ^2 \sin ^2( \omega_1 t)}{2 \omega_1^2}+1\right)\right] \\
    &~& + \frac{\beta k_B}{2\omega_1}\int_0^t T(s) e^{-\beta(t-s)}\left\{1+\beta\sin [2 \omega_1 (s-t)]+\frac{\beta^2 \sin ^2[\omega_1 (s-t)]}{\omega_1}\right\} \, ds. \nonumber
\end{eqnarray}
As one can see in Fig.~\ref{fig:Heat_avg}, our numerical results present a nice agreement with what is expected from the theory.
The system is continuously exchanging heat, following the thermal modulation $T(t)$. Comparing the green ($\Omega=1$) and orange ($\Omega=5$) curves in both graphs, we can see the effect of the driving frequency. For instance, for both the light (Fig.~\ref{fig:Heat_avg_m1}) and heavy (Fig.~\ref{fig:Heat_avg_m10}) particles, the slower driving frequency results in a much larger amplitude of heat exchange. This happens because  when the temperature changes slowly ($\Omega=1$), the system has more time to absorb or release heat before the cycle reverses. On the other hand, when the driving is fast ($\Omega=5$), the temperature changes so rapidly that the particle can't exchange much heat before it reverses, leading to much smaller oscillations in $\langle Q(t) \rangle$. Furthermore, since the thermal protocol modulates noise strength and therefore the statistical energy exchange, it couples strongly into the particle’s energetic degrees of freedom. This produces a large mean heat response for $\Omega$ near $\omega_0$: the green $m=1$ curve exhibits the largest amplitude — a resonant-like amplification.

By comparing Fig.~\ref{fig:Heat_avg_m1} ($m=1$) with Fig.~\ref{fig:Heat_avg_m10} ($m=10$), the role of inertia becomes clear.  The heavier particle has a significantly smaller amplitude of heat exchange compared to the lighter particle for the same driving frequency. This happens because inertia acts as a memory device and resists changes in motion: the heavier particle cannot respond as effectively to the changing thermal environment. Its kinetic energy lags behind the bath temperature, leading to less efficient and smaller heat exchange in each cycle.

\begin{figure}[ht!]
  \centering
  \begin{subfigure}[b]{0.32\linewidth}
    \subcaption{}     
    \includegraphics[width=\linewidth]{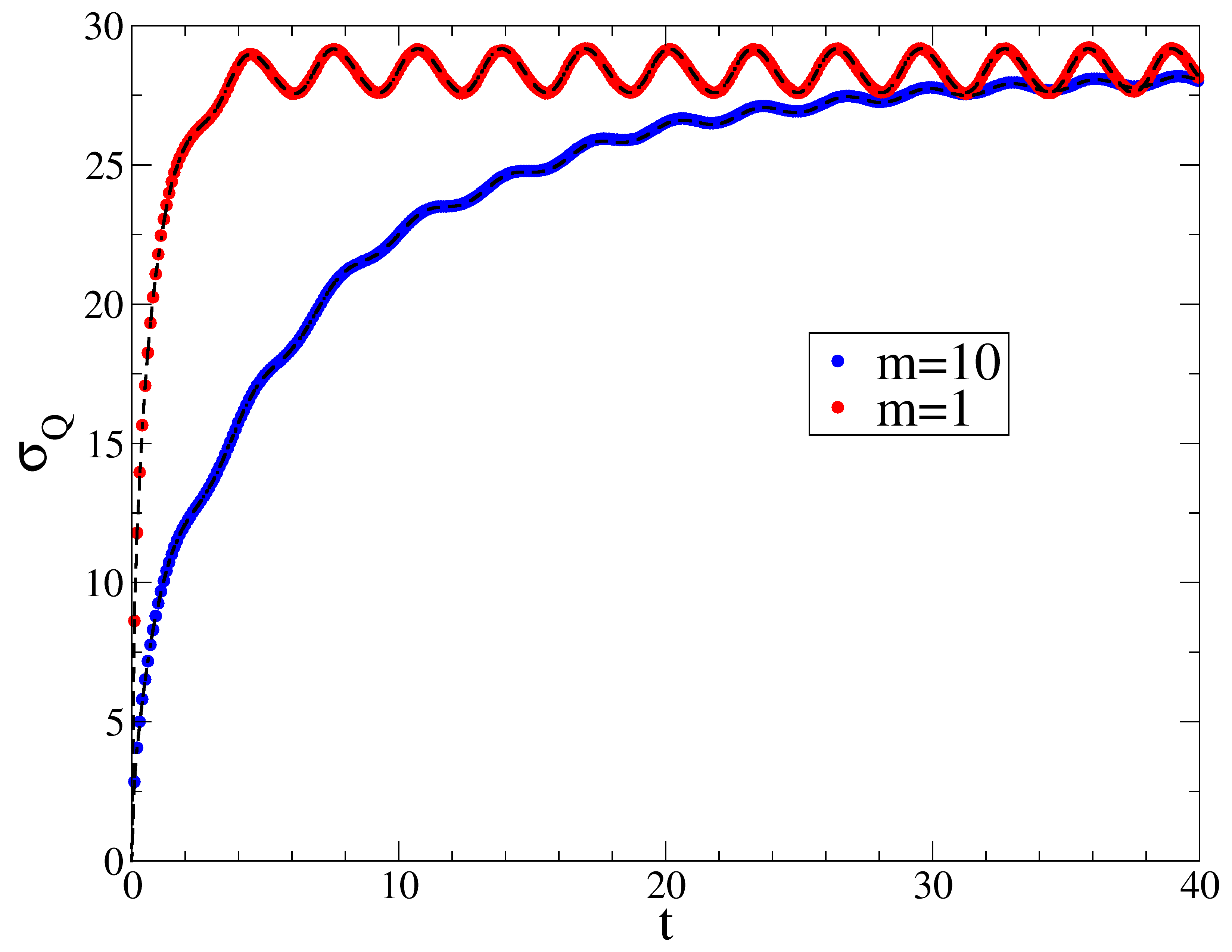}
    \label{fig:Heat_Stand_dev}
  \end{subfigure}  
  \begin{subfigure}[b]{0.32\linewidth}
    \subcaption{}       
    \includegraphics[width=\linewidth]{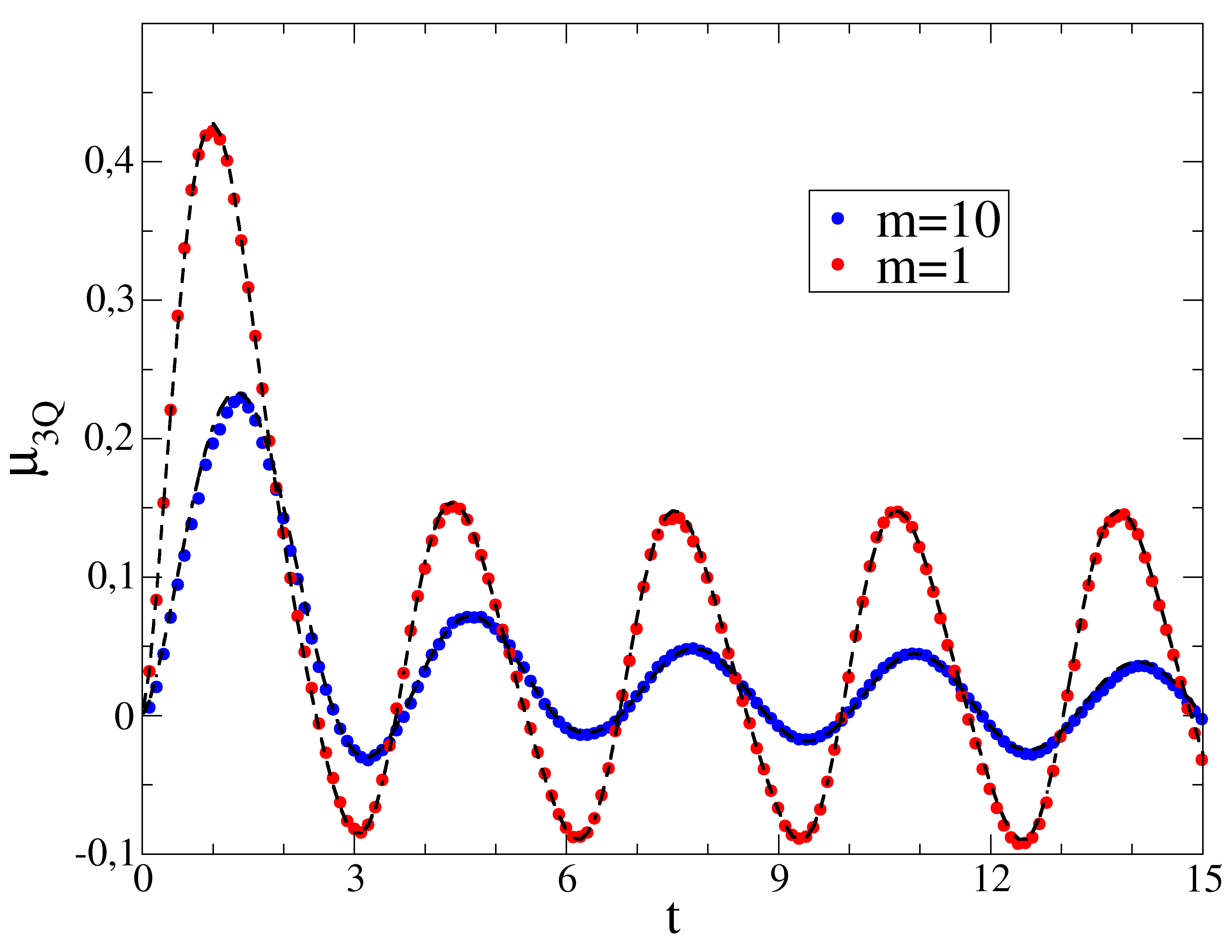}
    \label{fig:Heat_Skewness}
  \end{subfigure}
  \begin{subfigure}[b]{0.32\linewidth}
    \subcaption{}     
    \includegraphics[width=\linewidth]{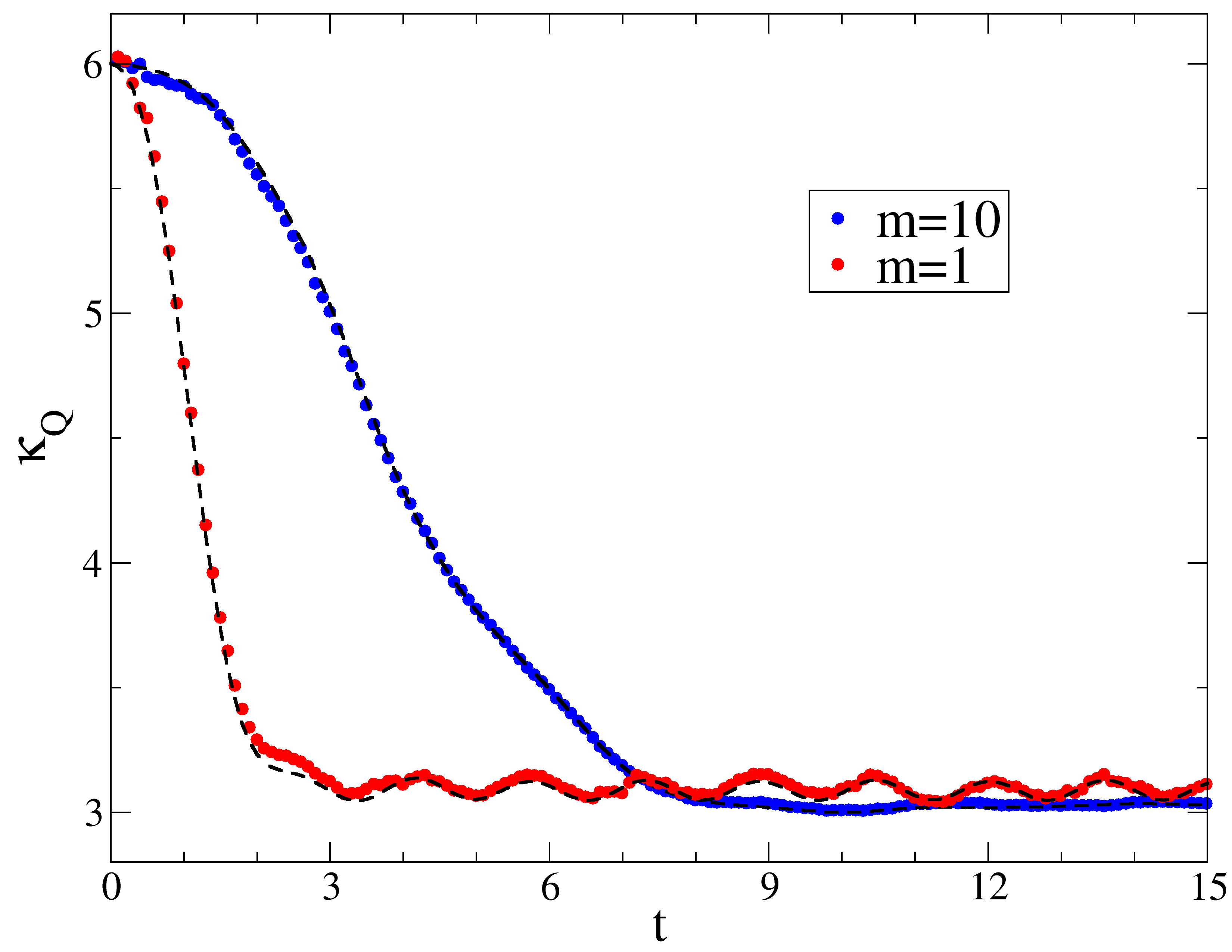}
    \label{fig:Heat_Kurtosis}
  \end{subfigure}
  \caption{(a) Heat standard deviation as function of time, (b) skewness as function of time, and (c) excess Kurtosis as function of time for the Brownian particle under the sinusoidal-temperature protocol. In both graphs, filled symbols correspond to data obtained from Brownian simulations with $k= \gamma= 1$, $T_0=20$, $T=5$ and $\Omega=2$, and the dashed lines are the results expected from theory.}
  \label{fig:Heat_statistics}
\end{figure}

The other statistical moments can be calculated by differentiating the characteristic function, Eq.\eref{Z_lambda}, so that
\begin{eqnarray}
    \langle Q^n\rangle = i^n  \frac{\partial^nZ(\lambda)}{\partial\lambda^n}\Big|_{\lambda=0}.
\end{eqnarray}
From the foregoing equation, we analyze the standard deviation, skewness, and excess kurtosis, as shown in Fig.~\ref{fig:Heat_statistics}.
Numerical results presented in Fig.~\ref{fig:Heat_Stand_dev} show that the light particle reaches the plateau much faster and has larger oscillations, whereas the heavy particle shows a slower rise and approaches more gradually. Fig.~\ref{fig:Heat_Skewness} confirms that the distribution remains asymmetric about its mean. In particular, the skewness $\mu_{3Q}$ changes sign periodically, and its positive peaks exceed the magnitude of its negative troughs. This bias toward positive skew reflects the fact that under the sinusoidal protocol, the particle absorbs heat more effectively than it releases. Moreover, an increase in the particle mass both reduces the amplitude of these skewness oscillations (damps the system's response) and shifts the oscillations in time (creates a phase lag). Finally, the excess kurtosis displayed in Fig.~\ref{fig:Heat_Kurtosis} starts at $\kappa_Q = 6$, indicating a strongly non-Gaussian, leptokurtic distribution and validating Eq.\eref{Heat_FT} at short times. As the system evolves, it relaxes into a periodic nonequilibrium steady state, set by the continuous driving. Furthermore, heavier particles, with greater inertia, exhibit longer relaxation times, so mass significantly prolongs the approach to that steady state.

\section{Conclusion}
\label{Conclusion}
In this work, we have studied a harmonically bound Brownian particle subjected to a sinusoidal thermal protocol. We employ the Langevin formalism to describe the dynamics and the stochastic energetics, which in this setup reduce to heat and internal energy. We analyze the Pearson correlation coefficient between displacement and velocity, and we derive analytical expressions for the characteristic function of the heat distribution from which its statistical moments are obtained. The heat distribution is evaluated numerically, and all theoretical predictions agree well with numerical simulations (see \ref{num_sim} for simulation details).

We find that the system dynamics and heat exchange depend sensitively on several parameters, including the driving frequency, the amplitude of the temperature oscillation, and the particle mass. The thermal frequency induces a time-scale competition between driving and relaxation: higher $\Omega$ would strengthen asymmetry (less time to equilibrate), while lower $\Omega$ would weaken it. This effect manifests in high order moments: both skewness and excess kurtosis are non-zero for $t>0$, confirming that the heat distribution $P(Q)$ is fundamentally non-Gaussian. Moreover, the heat flux is time-dependent, as expected, due to the thermal noise modulation. In the long-time limit, the driving periodicity is inherited by the probability distribution and by ensemble averages \cite{Reimann_1996,Morgado_2009,Chen_2023}; consequently, the system relaxes to a time-periodic non-equilibrium steady state.

Inertial effects — and thus the Pearson correlation coefficient — remain important when $T(t)$ changes on a time scale comparable to or faster than the velocity relaxation time. The particle's mass is a critical parameter; it dampens the system's response (lower amplitude of skewness oscillations), delays the response (phase lag in skewness) and slows down the relaxation to a steady state dramatically (longer transient time for standard deviation and excess kurtosis).

In conclusion, the results presented here demonstrate that dynamical and inertial contributions to the heat, induced by rapid thermal oscillations, produce complex hysteresis in energy flux. Furthermore, our findings indicate that time-periodic temperature modulations offer a viable route to control energy flow at the nanoscale. Exploiting these effects to design thermal devices constitutes a promising direction for future research.

\section*{Acknowledgments}
FPA would like to acknowledge Dr. Leandro G. Rizzi for insightful discussion. The authors are especially grateful to Dr. Pedro V. Paraguass\'{u}, whose feedback and extensive discussions were fundamental to this study. Although not listed as co-author, his contributions were essential to the development of this manuscript.

This work is supported by the Brazilian agencies Coordenação de Aperfeiçoamento de Pessoal de Ensino Superior (CAPES) and Fundação Carlos Chagas Filho de Apoio à Pesquisa do Estado do Rio de Janeiro (FAPERJ). FPA acknowledges CAPES scholarship, finance code 001 and FAPERJ  (Grants No. E-26/201.820/2025). WAMM acknowledges CNPq (Grant No. 308560/2022-1).

\appendix
\section{Moments derivation}
\label{moments_derivation}
For a Brownian particle subject to a harmonic potential, we define the friction rate $\beta = \gamma/m$, natural frequency $\omega_{0}^{2} = k/m$, and assume that the stochastic force satisfies the fluctuation–dissipation relation  $\langle A(t)\rangle=0$
$\langle A(t)A(t')\rangle= 2k_{\mathrm{B}}T(t)\beta\delta(t - t')/m$. Under these definitions, the Langevin equation (Eq.\eref{LE} in the main text) reads
\begin{eqnarray}
    \label{Langevin_eq}
    \ddot{x}(t) + \beta\dot{x}(t) + \omega_0^2 x(t) = A(t).    
\end{eqnarray}
From this equation, we seek to obtain the probability distribution $P(x,v,t;x_0,v_0,t_0)$. To do so, we first obtain the formal solution of Eq.\eref{Langevin_eq} regarded as an ordinary nonhomogeneous differential equation. The solution can be obtained in terms of the solution of the homogeneous equation given by 
\begin{eqnarray}
    \label{x_LE}
    x(t) = Ae^{\alpha_1 t} + Be^{\alpha_2 t},
\end{eqnarray}
where $\alpha_{1,2} = -\beta/2 \pm i\omega_1$ and $\omega_1 = \sqrt{\omega_0^2-\beta^2/4}$. To obtain the full inhomogeneous solution, we let the constants $A,B$ become time-dependent functions and impose the auxiliary condition~\cite{Chandrasekhar1943}
\begin{eqnarray}
    \label{LE_restriction1}
    \frac{dA}{dt}e^{\alpha_1 t} + \frac{dB}{dt}e^{\alpha_2 t} = 0,   
\end{eqnarray}
as well as the initial conditions $x(0) = x_0$ and $v(0) = v_0$. Therefore, one finds the solution
\begin{eqnarray}
   \label{x_t}
    x(t) &=& x_0 e^{-\beta t/2} \cos(\omega_1 t) + \frac{x_0 \beta + 2v_0}{2\omega_1} e^{-\beta t/2} \sin(\omega_1 t)  \\
    && + \frac{1}{\omega_1}\int_0^t A(s)e^{-\beta (t-s)/2} \sin[\omega_1 (t-s)] \,ds \nonumber\\ \nonumber \\
    \label{v_t}    
    v(t) &=& v_0 e^{-\beta t/2} \cos(\omega_1 t) - \frac{2x_0\omega_0^2 + \beta v_0}{2\omega_1}e^{-\beta t/2} \sin(\omega_1 t) \\ 
    && + \frac{1}{\omega_1}\int_0^t A(s)e^{-\beta (t-s)/2}\left\{ \omega_1 \cos[\omega_1 (t-s)] - \frac{\beta}{2}\sin[\omega_1 (t-s)] \right\}  \, ds.\nonumber
\end{eqnarray}
Imposing an initial equilibrium distribution at temperature $T_0$, \textit{i.e.}, $\langle x_0\rangle=\langle v_0\rangle=0$, $\langle x_0^2\rangle=k_B T_0/k$, and $\langle v_0^2\rangle=k_B T_0/m$ and using the fluctuation–dissipation relation, one shows from equations Eq.\eref{x_t} and Eq.\eref{v_t} that $\langle x(t)\rangle =\langle v(t)\rangle=0$, and the second moments and cross-correlation, respectively, reads
\begin{eqnarray}
    \langle x^2(t) \rangle &=& \frac{k_B T_0 e^{-\beta t} (\beta  \sin (\omega_1 t)+2 \omega_1 \cos (\omega_1 t))^2}{4 k \omega_1^2}+\frac{k_B T_0 e^{-\beta t} \sin ^2(\omega_1 t)}{m \omega_1^2} \\ && + \frac{2\beta k_B}{m\omega_1^2}\int_0^t T(s) e^{-\beta  (t-s)} \sin ^2[\omega_1(s-t)] \,  d s \nonumber\\
    \langle v^2(t) \rangle &=& \frac{k_B T_0 \omega_0^4 e^{-\beta t} \sin ^2(\omega_1 t)}{k \omega_1^2}+\frac{k_B T_0 e^{-\beta t} (\beta  \sin (\omega_1 t)-2 \omega_1 \cos (\omega_1 t))^2}{4 m \omega_1^2} \\ &&+ \frac{\beta k_B}{2m\omega_1^2}\int_0^t T(s) e^{-\beta(t-s)} \{\,\beta  \sin [\omega_1 (s-t)] + 2\omega_1 \cos [\omega_1 (s-t)]\,\}^2 \, d s \nonumber \\
    \langle x(t)v(t) \rangle &=& -\frac{\beta  k_B T_0 e^{-\beta t} \sin ^2(\omega_1 t)}{m \omega_1^2} \\ &&+ \frac{\beta k_B}{m\omega_1^2}\int_0^t  T(s)e^{-\beta(t-s)} \left\{\omega_1 \sin [2 \omega_1(t-s)]- \beta\sin ^2[\omega_1 (s-t)]\right\} d s. \nonumber
\end{eqnarray}
Initially, there was a Heaviside step function $\theta(t-s)$ in the integrand of the equations above, but the term can be suppressed since the integration limit is from $0$ to $t$ and in this region $\theta(t-s) =1$. Since $\langle x(t)\rangle =\langle v(t)\rangle=0$, the variances are $\sigma_x^2(t)=\langle x^2(t) \rangle$ and $\sigma_v^2(t)=\langle v^2(t) \rangle$. After some further manipulation on the foregoing expressions, equations Eq.\eref{x2_med}, Eq.\eref{v2_med} and Eq.\eref{xv_med} in the main text are obtained.

\subsection{Moments for the sinusoidal thermal protocol}
Considering the sinusoidal temperature protocol $T(t) = T_0 + T\sin(\Omega t)$ and some further manipulation, equations Eq.\eref{x2_med}, Eq.\eref{v2_med} and Eq.\eref{xv_med} in the main text reads
\begin{eqnarray}
    \sigma_x^2(t) &=& \frac{k_B T_0}{k} + \frac{4 \beta  k_B T \left[\Omega  \left(\Omega ^2 -3 \beta ^2 -4 \omega_1^2\right) \cos(\Omega t)+\beta  \left(4 \omega_0^2-3 \Omega ^2\right) \sin(\Omega t)\right]}{m \left(\beta ^2+\Omega ^2\right) \left(4 \beta ^2 \Omega ^2+\left(\Omega ^2-4 \omega_0^2\right)^2\right)} \\ 
      &&+\frac{\beta  k_B T \Omega  e^{-\beta t}}{m \omega_1^2} \left(\frac{4 \beta  \omega_1 \sin(2\omega_1 t)- \left[\Omega ^2+4 \omega_0^2-8 \omega_1^2\right]\cos(2\omega_1 t)}{4 \beta ^2 \Omega ^2+\left(\Omega ^2-4 \omega_0^2\right)^2} + \frac{1}{\beta ^2+\Omega ^2}\right)  \nonumber \\ \nonumber \\
    \sigma_v^2(t) &=& \frac{k_B T_0}{m} + \frac{\beta  k_B T \left[\Omega ^3 \left(\beta ^2+12 \omega_1^2\right)-2 \Omega ^5-16 \Omega  \omega_0^4\right]\cos(\Omega t)}{m \left(\beta ^2+\Omega ^2\right) \left(4 \beta ^2 \Omega ^2+\left(\Omega ^2-4 \omega_0^2\right)^2\right)} \\
      && + \frac{\beta^2 k_B T \left[\Omega ^2 \left(\beta ^2-12 \omega_1^2\right)+4 \Omega ^4+16 \omega_0^4\right]\sin(\Omega t)}{m \left(\beta ^2+\Omega ^2\right) \left(4 \beta ^2 \Omega ^2+\left(\Omega ^2-4 \omega_0^2\right)^2\right)} - \frac{\beta  k_B T \Omega  e^{-\beta t}}{m \omega_1^2}\left(\frac{\omega_0^2}{\beta ^2+\Omega ^2} \right) \nonumber \\
      && + \frac{\beta  k_B T \Omega  e^{-\beta t}}{m \omega_1^2}\left( \frac{ 4 \beta  \Omega ^2 \omega_1 \sin(2\omega_1 t) + \left[\Omega ^2(\beta ^2 -4\omega_1^2)+16 \omega_0^4\right]\cos(2\omega_1 t) }{4 \left(4 \beta ^2 \Omega ^2+\left(\Omega ^2-4 \omega_0^2\right)^2\right)} \right) \nonumber \\
      \nonumber \\
    \langle x(t)v(t) \rangle &=& -\frac{2 \beta  k_B T \Omega  \left[\Omega  \left(\Omega ^2 -3 \beta ^2-4 \omega_1^2\right) \sin(\Omega t)-\beta  \left(4 \omega_0^2 -3 \Omega ^2\right) \cos(\Omega t)\right]}{m \left(\beta ^2+\Omega ^2\right) \left(4 \beta ^2 \Omega ^2+\left(\Omega ^2-4 \omega_0^2\right)^2\right)} \\
      &&+ \frac{\beta  k_B T \Omega  e^{-\beta t}}{2 m \omega_1^2} \left(\frac{2 \omega_1 \left(\Omega ^2-4 \omega_0^2\right) \sin(2\omega_1 t) + \beta  \left(\Omega ^2+4 \omega_0^2\right) \cos(2\omega_1 t)}{4 \beta ^2 \Omega ^2+\left(\Omega ^2-4 \omega_0^2\right)^2} -\frac{\beta }{\beta ^2+\Omega ^2}\right) \nonumber
\end{eqnarray}

\section{Characteristic function}
\label{characteristic_func}

The evolution of the probability distribution $P(x,v,t)$ from $t=0$ to $\tau$  is described by the Klein-Kramers equation~\cite{Risken_1989}
\begin{equation}
    \label{Kramers}
    \partial_tP(x,v,t)
    =\left[
    - v\partial_x 
    + \partial_v \left( \gamma v + \frac{\partial_xV(x)}{m} \right)
    + \frac{\gamma k_B T(t)}{m} \partial_v^2
    \right] P(x,v,t).
\end{equation}
The Green's function of Eq.\eref{Kramers} can be written as the bivariate Gaussian distribution
\begin{eqnarray}
    P(x,v,t) &=& \frac{1}{2\pi\sigma_x(t)\sigma_v(t)\sqrt{1-\rho_{xv}^2(t)}} \\ && \exp\left(-\frac{1}{2[1-\rho_{xv}^2(t)]}\left[\left(\frac{x}{\sigma_x(t)}\right)^2 - 2\rho_{xv}\left(\frac{x}{\sigma_x(t)}\right)\left(\frac{v}{\sigma_v(t)}\right) + \left(\frac{v}{\sigma_v(t)}\right)^2\right]\right). \nonumber
\end{eqnarray}
If the initial probability distribution is Gaussian and is uncorrelated in position and velocity, \textit{i.e.},
\begin{eqnarray}
    P(x_0,v_0,0) = \frac{1}{2\pi\sigma_{x_0}\sigma_{v_0}}\exp\left(-\frac{x_0^2}{\sigma_{x_0}}-\frac{v_0^2}{\sigma_{v_0}}\right),
\end{eqnarray}
then the distribution remains Gaussian for all later times and is given by
\begin{eqnarray}    
    \label{Prob_dist}
    P(x,v,x_0,v_0) = \frac{1}{\sqrt{(2\pi)^4\det \mathbf{\Sigma}}} 
\exp\left(-\frac{1}{2}\mathbf{z}^T\mathbf{\Sigma}^{-1}\mathbf{z} \right),
\end{eqnarray}
where $\mathbf{z} = [x,x_0,v,v_0]^T$ is a $4$-dimensional column vector, and $\mathbf{\Sigma}$ is the covariance matrix 
\begin{eqnarray}
    \mathbf{\Sigma} = \left[ \begin{array}{cccc}
        \sigma_x^2(t) & \langle x(t)x_0\rangle & \langle x(t)v(t)\rangle & \langle x(t)v_0\rangle \\ \langle x(t)x_0\rangle & \sigma_{x_0}^2 & \langle v(t)x_0\rangle & \langle x_0v_0\rangle \\ \langle x(t)v(t)\rangle & \langle v(t)x_0\rangle & \sigma_v^2(t) & \langle v(t)v_0\rangle \\ \langle x(t)v_0\rangle & \langle x_0v_0\rangle & \langle v(t)v_0\rangle & \sigma_{v_0}^2
    \end{array}\right].
\end{eqnarray}
Its elements are obtained using Eq.\eref{x_t} and Eq.\eref{v_t} and the initial equilibrium distribution, such that
\begin{eqnarray}
    \langle x_0v_0\rangle &=& 0 \\
    \sigma_{x_0}^2 &=& \frac{k_BT_0}{k} \\
    \sigma_{v_0}^2 &=& \frac{k_BT_0}{m} \\
    \langle x(t)x_0\rangle &=& \frac{k_B T_0 e^{-\beta  t/2}}{2 k \omega_1}[2 \omega_1 \cos(\omega_1 t)+\beta  \sin(\omega_1 t)] \\
    \langle x(t)v_0\rangle &=& \frac{k_B T_0 e^{-\beta  t/2}}{m \omega_1}\sin(\omega_1 t) \\ 
    \langle v(t)x_0\rangle &=& -\frac{k_B T_0 e^{-\beta  t/2}}{m \omega_1}\sin(\omega_1 t)\\   
    \langle v(t)v_0\rangle &=& \frac{k_B T_0 e^{-\beta  t/2}}{2 m \omega_1}[2 \omega_1 \cos(\omega_1 t)-\beta  \sin(\omega_1 t)].  
\end{eqnarray}
The characteristic function can be obtained by applying Eq.\eref{Prob_dist} and Eq.\eref{stc_Q} in Eq.\eref{Z_func} and performing the Gaussian integrals, such that 
\begin{eqnarray}
    Z(\lambda) &=& \frac{1}{\sqrt{(2\pi)^4\det \mathbf{\Sigma}}} \int dx \int dv \int dx_0 \int dv_0 ~ \exp\left(-\frac{1}{2}\mathbf{z}^T\mathbf{M}\mathbf{z} \right) \nonumber \\
    &=&  \frac{1}{\sqrt{(2\pi)^4\det \mathbf{\Sigma}}}\sqrt{\frac{(2\pi)^4}{\det \mathbf{M}}} = \frac{1}{\sqrt{\det\mathbf{\Sigma}\cdot\det\mathbf{M}}},
\end{eqnarray}
where $\mathbf{M} = \mathbf{Q}+ \mathbf{\Sigma}^{-1}$ and
\begin{eqnarray}
    \mathbf{Q} = i\lambda\left[ \begin{array}{cccc}
        k & 0 & 0 & 0 \\
        0 & -k & 0 & 0 \\
        0 & 0 & m & 0 \\
        0 & 0 & 0 & -m 
    \end{array}\right].
\end{eqnarray}
Finally, after further simplification, the characteristic function becomes
\begin{eqnarray}
    Z(\lambda) = \frac{1}{\sqrt{\det(\mathbf{\Psi}+\mathbf{I}_4)}}
\end{eqnarray}
where $\mathbf{I}_4$ is the identity matrix and
\begin{eqnarray}
    \mathbf{\Psi} = i\lambda\left[ \begin{array}{cccc}
        k\, \sigma_x^2 & k\,\langle x(t)x_0\rangle & k\,\langle x(t)v(t)\rangle & k\,\langle x(t)v_0\rangle \\ -k\,\langle x(t)x_0\rangle & -k\,\sigma_{x_0}^2 & -k\,\langle v(t)x_0\rangle & 0 \\ m\,\langle x(t)v(t)\rangle & m\,\langle v(t)x_0\rangle & m\,\sigma_v^2 & m\,\langle v(t)v_0\rangle \\ -m\,\langle x(t)v_0\rangle & 0 & -m\,\langle v(t)v_0\rangle & -m\,\sigma_{v_0}^2
    \end{array}\right].     
\end{eqnarray}
The probability normalization condition imposes that $Z(0)=1$.

\section{Numerical simulations}
\label{num_sim}
Considering a finite time interval $\Delta t= t_{i+1} - t_i$, Eq.\eref{LE} can be discretized and rewritten as 
\begin{eqnarray}
    \label{disc_traj}
    x_{i+1} = x_i + v_i \Delta t, \quad v_{i+1}=v_i - \frac{\gamma}{m} v_i \Delta t - \frac{k}{m}x_i\Delta t + \tilde{F}_a\Delta t.
\end{eqnarray}
$\tilde{F}_a(t)$ is a random force evaluated as~\cite{Azevedo2020}
\begin{eqnarray}
    \tilde{F}_a(t) = \sqrt{\frac{2\gamma k_B T(t)}{m^2 \Delta t }} N(0,1),    
\end{eqnarray}
where $N(0,1)$ denotes a Gaussian variable with zero mean and variance equal to one. Thus, the stochastic heat can be obtained in terms of the trajectory, Eq.\eref{disc_traj}, such that 
\begin{eqnarray}
    Q_i = \frac{1}{2}m(v_{i}^2 - v_{i-1}^2) + \frac{1}{2}k(x_{i}^2 - x_{i-1}^2).   
\end{eqnarray}

This method is a physical approach provided by a simple discrete-state jump Markov process; it allows modeling the immediate effects of individual molecular collisions on the velocity of the Brownian particle. Despite this being a discrete and idealized model, the continuum limit duplicates the predictions of the Langevin equation~\cite{Gillespie1993}.

For each simulation we consider $N_t=10^6$ independent trajectories, a time interval $\Delta t = 10^{-4}$, effective spring constant $k=1$, temperatures $T_0=20$ and $T=5$, and the Boltzmann constant $k_B=1$. Also, we set initial conditions $x_0$ and $v_0$ in thermal equilibrium at temperature $T_0$. So, for each trajectory 
\begin{eqnarray}
    x_0 = \frac{k_B T_0}{k}N(0,1), \quad v_0= \frac{k_B T_0}{m}N(0,1). 
\end{eqnarray}

\providecommand{\newblock}{}

\end{document}